\documentclass[manuscript]{aastex}

\usepackage{subfigure}

\slugcomment{to appear in the {\it Astronomical Journal}}

\shorttitle{Parallaxes from CTIOPI}
\shortauthors{Jao et al.}

\begin{document}

\title{The Solar Neighborhood XXXI: Discovery of an Unusual Red+White
  Dwarf Binary at $\sim$25 Parsecs via Astrometry and UV Imaging}

\author{Wei-Chun Jao\altaffilmark{1}, Todd J. Henry\altaffilmark{1}}
\affil{Department of Physics and Astronomy, Georgia State University,
 Atlanta, GA 30302}
\email{jao@chara.gsu.edu, thenry@chara.gsu.edu}

\author{John P. Subasavage\altaffilmark{1}}
\affil{US Naval Observatory, Flagstaff Station, 10391 West
Naval Observatory Road, Flagstaff, AZ 86001}
\email{jsubasavage@nofs.navy.mil}  

\author{Jennifer G. Winters\altaffilmark{1}, Douglas R. Gies}

\affil{Department of Physics and Astronomy, Georgia State University,
Atlanta, GA 30302}
\email{winters@chara.gsu.edu, gies@chara.gsu.edu}

\author{Adric R. Riedel\altaffilmark{1}}
\affil{Department of Physics and Astronomy, Hunter College, 695 Park Avenue, New York, NY 10065}
\email{ar494@hunter.cuny.edu}

\and 

\author{Philip A. Ianna\altaffilmark{1}}
\affil{Department of Astronomy, University of Virginia,
Charlottesville, VA 22904-4325}
\email{philianna3@gmail.com}

\altaffiltext{1}{Visiting Astronomer, Cerro Tololo Inter-American
Observatory.  CTIO is operated by AURA, Inc.\ under contract to the
National Science Foundation.}

\begin{abstract}

We report the discovery of a nearby M5.0V dwarf at 24.6 pc, SCR
1848$-$6855, that is orbited by an unusual companion causing an
astrometric perturbation of more than 200 milli-arcseconds.  This is
by far the largest perturbation found to date among more than 700
targets observed during our long-term astrometry/photometry program at
the CTIO 0.9-m telescope.  We present here a suite of astrometric,
photometric, and spectroscopic observations of this high proper motion
($\sim$1\farcs3/yr) system in an effort to reveal the nature of this
unusual binary.  The measured near-UV and optical U band fluxes exceed
those expected for comparable M5.0V stars, and excess flux is also
detected in the spectral range 4000\AA--7000\AA.  The elusive
companion has been detected in HST-STIS$+$MAMA images at 1820\AA ~and
2700\AA, and our analysis shows that it is probably a rare, cool,
white dwarf with T = 4600-5500K.  Given the long-term astrometric
coverage, the prospects for an accurate mass determination are
excellent, although as yet we can only provide limits on the unusual
companion's mass.

\end{abstract}


\section{Introduction}

Using the CTIO 0.9-m telescope, we have made astrometric observations
of more than 700 nearby stars since 1999 during our Cerro Tololo
Inter-american Observatory Parallax Investigation (CTIOPI).  We have
presented nearby subdwarf, red dwarf, and white dwarf discoveries in
{\it The Solar Neighborhood} series \citep{Jao2005, Henry2006,
Subasavage2009, Riedel2010, Jao2011, Riedel2011}, as part of the
RECONS (REsearch Consortium On Nearby Stars)\footnote{www.recons.org}
effort to explore the solar neighborhood.  To date, we have reported
astrometric perturbations with amplitudes of tens of milli-arcseconds
(hereafter, mas) \citep{Henry2006, Riedel2010} and continue to observe
$\sim$50 objects exhibiting perturbations in an effort to determine
accurate masses for low mass stars and brown dwarfs.  Here we report
an extraordinary perturbation of the M5.0V star SCR 1848-6855
(hereafter SCR 1848) with an amplitude of more than 200 mas.  This
star was discovered during our SuperCOSMOS-RECONS (SCR) search for
high proper motion objects in the southern sky and found to have
$\mu=$1\farcs287 yr$^{-1}$ \citep{Hambly2004}.  In this paper, we
present a detailed analysis of this unusual system, including results
from the 0.9-m program and several other ground-based and space-based
efforts, to understand the physical properties of the mysterious
companion.

\section{Observations and Results}

\subsection{Astrometry from CTIO}
\label{sec:astrometry}

We used the CTIO 0.9-m to measure the astrometric properties (proper
motion, parallax, perturbation of the photocenter) and $UBVRI$
photometry of SCR 1848, and provide the results in
Table~\ref{tbl:basic}.  The telescope has a 2048$\times$2048 Tektronix
CCD camera with 0\farcs401 pixel$^{-1}$ plate scale \citep{Jao2005}.
For both astrometric and photometric observations, we used the central
quarter of the chip, yielding a 6\farcm8 square field of view.  We
used the Kron-Cousins $I$ filter for parallax measurements to maximize
the number of suitable reference stars in the field, and obtained
$UBVRI$ photometry of SCR 1848 and similar stars through Johnson ($U$,
$B$ and $V$) and Kron-Cousins ($R$ and $I$) filters.  Bias and dome
flat frames were taken nightly for basic image reduction.  We have one
epoch of $U$ and $B$ photometry and four epochs of $V$, $R$ and $I$,
for which results are given in Table~\ref{tbl:basic}.  Additional
details of parallax and photometry observations and data analysis can
be found in \cite{Jao2005} and \cite{Winters2011}, respectively.

\begin{figure}
  \vspace*{-2cm}
  \centering
  
  \subfigure[SCR1848]{
  \includegraphics[scale=0.34, angle=90]{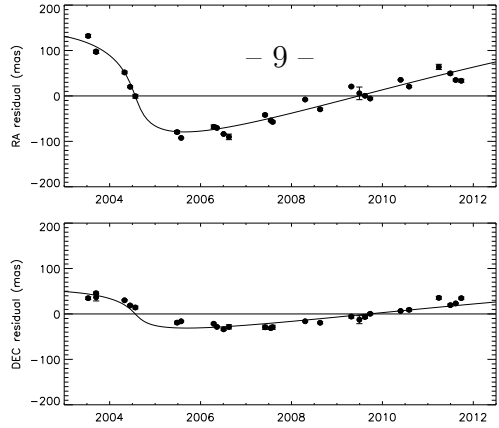}
  \label{fig:1a}
  }
  
  \subfigure[2MASS J18482451-6854429]{
  \includegraphics[scale=0.34, angle=90]{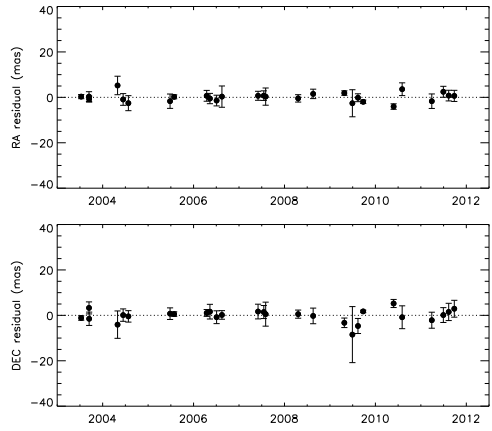}
  \label{fig:1b}
  }
  
  \subfigure[SCR 1848 between 2007 and 2012]{
  \includegraphics[scale=0.34, angle=90]{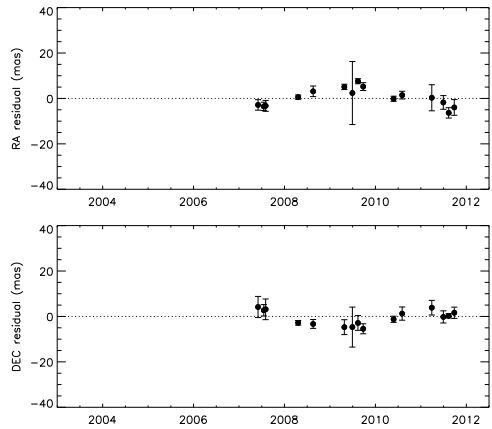}
  \label{fig:1c}
  }
  
 \caption {Figure 1(a): Astrometric residuals in the RA and DEC
   directions are shown for SCR 1848 after solving for proper motion
   and parallax.  Each point represents typically 5--10 frames taken
   on a single night.  The line traces a fit to the photocenter
   shift, i.e., the perturbation, as discussed in
   section~\ref{sec:companion}.  Figure 1(b): Residuals for a
   reference star, 2MASS J18482451-6854429, in the same field are
   shown for comparison.  Figure 1(c): The residuals of SCR 1848
   using data from 2007-2012 only are shown, avoiding the change in
   the perturbation trend around 2005.  Note that the scales in the
   vertical axes of b and c are a factor of five less than in
   a. \tiny{(NOTE TO EDITOR: PLEASE ARRANGE THIS FIGURE FROM LEFT TO RIGHT,
   TWO PANELS EACH.)}}
  
  \label{fig:perturb}
  \end{figure}

With more than 8 years of astrometric observations, the red dwarf
exhibits an extraordinary perturbation with peak-to-peak amplitudes of
$\sim$200 mas in RA and $\sim$60 mas in DEC.  Shown in
Figure~\ref{fig:1a} is a map of the red dwarf's photocentric position
relative to the barycenter, split into RA and DEC directions, after
shifts due to proper motion and parallax have been removed.  We
suspect that the epoch of periastron occurred around 2004.6.  This is
the largest photocentric perturbation we have detected among more than
700 targets observed astrometrically at the 0.9-m since 1999.  To
eliminate any doubt that the perturbation is real, for comparison we
also present residuals from a reference star in the field in
Figure~\ref{fig:1b}, in which the vertical scale is one-fifth that for
SCR 1848.

Because the scale of the perturbation exceeds the size of the
parallactic ellipse by a factor of five (so far), the parallactic
motion is overwhelmed by the perturbation, and the yearly cadence of
the parallax remains in Figure~\ref{fig:1a}, particularly in the RA
axis.  To derive an accurate trigonometric parallax for the system, we
therefore solve for proper motion, parallax, and orbital motion
iteratively.  We first carry out a standard reduction for proper
motion and parallax. A reduction using all of the data without
compensating for the perturbation results in a parallax of
24.26$\pm$2.30 mas (41 pc), which is inconsistent with the more
carefully treated results discussed below. We then fit an photocentric
orbit to the residuals shown in Figure~\ref{fig:1a}. This orbit is
then removed, and we derive the results for proper motion
(1280.9$\pm$0.3 mas/yr) and parallax (40.63$\pm$0.72 mas), placing the
system at a distance of 24.6 pc. After the first iteration of removing
the large perturbation, the residuals in RA were reduced by a factor
of 4.5 and the parallax error was reduced by a factor of 3.  These new
residuals are comparable to the night-to-night variations in the
photocenter positions, and the parallax error is less than our median
parallax errors ($\sim$1.4 mas) for stars with coverage of at least
six years, so no further iterations were appropriate for the current
dataset. We will continue to follow the system to improve our
understanding of the overall path of the photocenter. Further
discussion is based upon the parallax of 40.63 mas as given in
Table~\ref{tbl:basic}. We checked this result by evaluating only those
images taken since 2007 to ensure that the perturbation persists,
i.e.,~ selecting frames after the presumed periastron, with results
shown in Figure~\ref{fig:1c}.  A comparison of the reference star's
motion in Figure~\ref{fig:1b} to the residual effect of the
perturbation in SCR 1848's position in Figure~\ref{fig:1c} (same scale
in both panels) reveals that the perturbation's effects are seen on
both axes and that the science star is not single.  The parallax
(46.26$\pm$0.70 mas) derived for the 2007-2012 data is roughly in
agreement with the results for all data with the perturbation removed
,although we note that in this unusual case the parallax errors may be
underestimated and the distance from 2007-2012 data should be treated
with caution\footnote{Because the perturbation in Figure~\ref{fig:1c}
is significantly reduced compared to that in Figure~\ref{fig:1a}, we
did not perform an orbital fit to revise the parallax.  We merely use
this reduction as additional assurance that the system is at a
distance of roughly 24.6 pc.}. Details pertaining to the perturbation
fit as well as the analysis of the photocentric orbit are discussed in
section~\ref{sec:companion}.


\begin{deluxetable}{lr}
\tablewidth{350pt}
\tabletypesize{\small}
\tablecaption{Astrometric and photometric results for SCR1848\label{tbl:basic}}
\tablehead{
\colhead{Result}  &
\colhead{Value}
}
\startdata
Name                   &        SCR 1848$-$6855       \\
R.A.(2000.0)           &        18:48:21.1           \\
Decl. (2000.0)         &     $-$68:55:34.5           \\
observation span(yr)   &        8.5                   \\
No. of reference stars &        11                    \\
No. of frames          &        141                   \\
$\pi_{rel}$(mas) &        39.52$\pm$0.72    \\
correction to absolute parallax (mas) &     1.11$\pm$0.08   \\       
$\pi_{abs}$(mas) &        40.63$\pm$0.72    \\
$\mu$(mas yr$^{-1}$) &      1280.9$\pm$0.3   \\
P.A. (degrees)          &        195.7$\pm$0.1      \\
$NUV_{2271}$           &        22.55               \\
$NUV_{2711}$           &        (23.60)             \\
$U$                    &        18.35               \\
$B$                    &        18.13               \\
$V$                    &        16.86               \\
$R$                    &        15.68               \\
$I$                    &        13.83               \\
$J$                    &        11.89               \\
$H$                    &        11.40               \\
$K_{s}$                &        11.10               \\
spectral type          &        M5.0VJ               
\enddata

\tablecomments{The $NUV_{2271}$ photometry is from GALEX General
  Release 6.  $NUV_{2711}$ is specifically of the M dwarf component
  measured directly from STIS images discussed in
  section~\ref{subsec:STISNUV}, but all other magnitudes are combined
  photometry of both components.  $U$ and $B$ were taken on March 12,
  2010 at the CTIO 0.9-m, while the $V$, $R$, and $I$ measurements are
  from four different nights of observations. $JHK_{s}$ photometry is
  from the 2MASS catalog. The ``J'' in the spectral type indicates it
  is classified from a combined spectrum of two components.}

%
%
%
\end{deluxetable}

    
\clearpage

\subsubsection{A Background Source Near SCR 1848}
\label{sec:background}
SCR 1848 is a high proper motion star that moves primarily south with
a slight deviation to the west.  After carefully checking all parallax
frames, we noticed it was on top of a faint background source in the
first image taken on UT 2003 March 29 by comparing it with the
SuperCOSMOS infrared plate taken in 1996.  The background star later
appeared clearly in our image on UT 2005 July 29 and in subsequent
frames.  During this period of time, SCR 1848 coincidentally shows its
largest parallax residual in RA, as shown in Figure~\ref{fig:1a}.  In
order to understand how this faint background star affects the
centroids by SExtractor, we performed the following simulations to
confirm that the perturbation is {\it not} due to this faint
background star.

First, we used the last few epochs of images, in which SCR 1848 and
the background source were widely separated (e.g.~11\farcs7 at
position angle 18$^{\circ}$ on UT 2011 September 26), to measure the
flux ratio of the two sources to be 400:1 in the $I$ filter used for
the parallax series.  No proper motion was detected for the background
source during the last two years of images, so we then created
simulated images of SCR 1848 and the background source by using PSFs
of reference stars in the SCR 1848 field utilizing the first image
taken in each of the 28 epochs, and adding in the background source at
its position measured at the last epoch.  We then created FITS images
containing the two sources at epochs corresponding to our parallax
frames by tracing SCR 1848's position from 2011 back to 2003 using SCR
1848's presumably uncorrupted proper motion and position angle values
from \cite{Hambly2004}.

Using SExtractor to determine the centroids of SCR 1848 in the
simulated images and in the original parallax frames, we find mean
differences of 4.7 and 12.9 mas in RA and DEC, respectively, in the 28
epochs of images.  Since 2005, the differences are less than 1 mas.
These results illustrate that the early epoch images have centroids
that are slightly skewed by the background star and that the offsets
are larger in DEC than in RA, which is contradictory to what we see in
the parallax residuals.  In fact, the offsets in RA are $\sim$40 times
smaller than what is measured for the perturbation shown in
Figure~\ref{fig:1a}.  We therefore conclude that the faint
background source is {\it not} the cause of the perturbation.

\subsection{$UBVRI$ Photometry from CTIO}

\begin{figure}
  \centering
  \includegraphics[angle=90, scale=0.7]{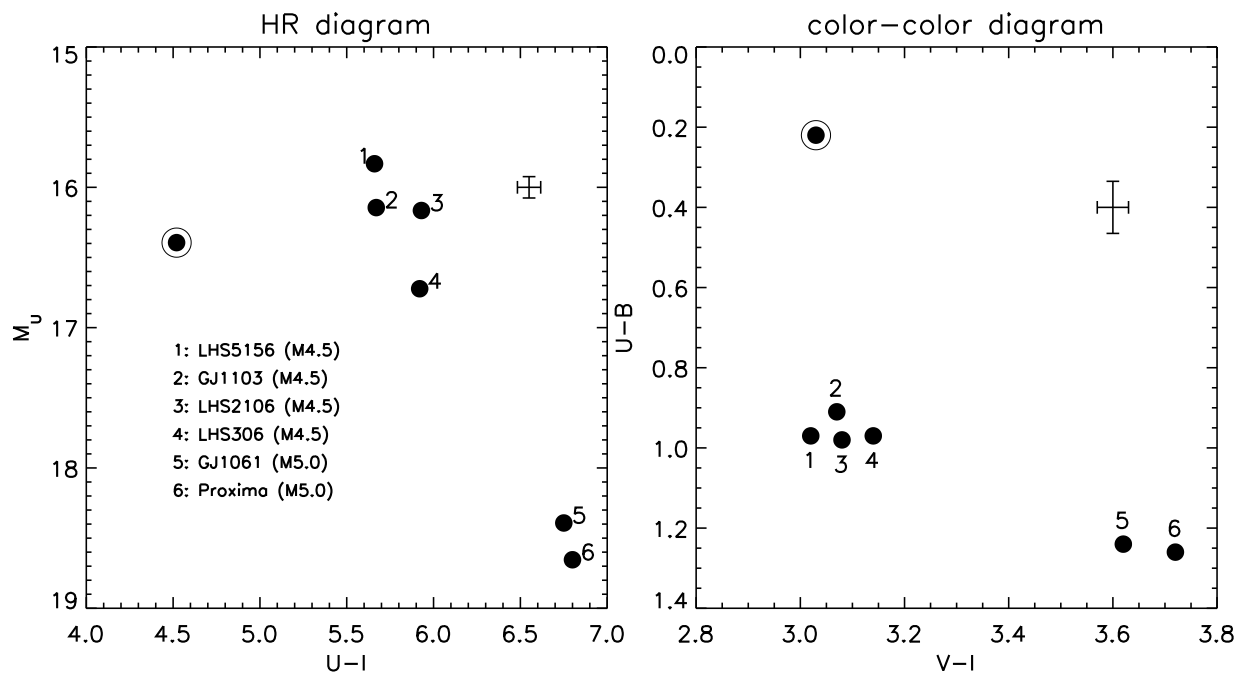}

  \caption{A color-absolute magnitude diagram (left) and a color-color
    diagram (right) show that SCR 1848 (circled points) is brighter
    and bluer than two comparable M5.0V spectroscopic standard stars,
    and does not match earlier type stars.  The broad band $UBVRI$
    photometry of these comparison targets and SCR 1848 was acquired
    at the CTIO 0.9-m using the instrumental setup discussed in
    section~\ref{sec:astrometry}.  Parallaxes of these comparison
    targets are from \cite{YPC} and RECONS.  The mean error bars on
    each axis are shown.}

  \label{fig:colorcolor}
  \end{figure}

The photometric distance estimated from $VRIJHK_{s}$ colors using the
suite of relations in \citep{Henry2004} is 37.0$\pm$9.4 pc, which is
1.3$\sigma$ further than the adopted trigonometric parallax distance
of 24.6 pc, but is not terribly useful as a distance constraint.  The
large error in the photometric estimate is caused by photometric
colors for the system that differ from similar stars.  $UBVRI$
photometry for SCR 1848 and six similar red dwarfs was obtained at the
CTIO 0.9-m on 2010 March 13, and is given in Table~\ref{tbl:UBVRI}.
As shown in Figure~\ref{fig:colorcolor}, in a $U-I$ color plot SCR
1848 is noticeably bluer than stars of comparable spectral types
spanning similar $M_U$, and does not follow the trend of similar red
dwarfs in the $U-B$ vs. $V-I$ color-color plot.  As discussed in
section~\ref{sec:redblue}, SCR 1848's red spectrum is a good match to
the spectroscopic standards Proxima and GJ1061 at type M5.0V, but the
$U$ and $B$ photometry indicates that the companion is contributing
blue light to the system.  If SCR 1848 had $M_{U}\sim$18.5, consistent
with other M5.0V, it would be only $\sim$9.3 pc away, which is
inconsistent with the trigonometric parallax.


\begin{deluxetable}{lccccccrll}
\tablecaption{$UBVRIK_{s}$of M dwarfs\label{tbl:UBVRI}}
\tablehead{
\colhead{Object}    &
\colhead{$U$} &
\colhead{$B$} &
\colhead{$V$} &
\colhead{$R$} &
\colhead{$I$} &
\colhead{$K_{s}$} &
\colhead{$\pi$} &
\colhead{SpecType}  &
\colhead{Ref} \\
\colhead{}    &
\colhead{} &
\colhead{} &
\colhead{} &
\colhead{} &
\colhead{} &
\colhead{} &
\colhead{(mas)} &
\colhead{}  &
\colhead{} 
}
\startdata
LHS5156 & 15.94 &  14.97  &   13.30 &  11.98 &   10.28 &  7.77 &  95.15$\pm$1.17  & M4.5V  & 1, 2, 7 \\
GJ1103  & 15.86 &  14.95  &   13.26 &  11.89 &   10.19 &  7.66 & 114.00$\pm$3.30  & M4.5V  & 1, 3, 7, 9\\
LHS2106 & 17.06 &  16.08  &   14.21 &  12.87 &   11.13 &  8.65 &  66.23$\pm$1.16  & M4.5V  & 1, 2, 7, 8\\
LHS306  & 16.97 &  16.00  &   14.19 &  12.81 &   11.05 &  8.50 &  89.24$\pm$1.69  & M4.5V  & 1, 2, 5, 6\\
GJ1061  & 16.22 &  14.98  &   13.07 &  11.45 &    9.47 &  6.61 & 271.92$\pm$1.34  & M5.0V  & 1, 2, 4\\
Proxima & 14.21 &  12.95  &   11.13 &   9.45 &    7.41 &  4.38 & 774.25$\pm$2.08  & M5.0V  & 1, 2, 6\\
SCR1848 & 18.35 &  18.13  &   16.86 &  15.68 &   13.83 & 11.10 &  40.63$\pm$0.72  & M5.0VJ & 1, 2
\enddata
\tablerefs{(1) this work;
(2) \cite{2mass};
(3) \cite{Deterich2012}; 
(4) \cite{Henry2006};
(5) \cite{Hawley1996};
(6) \cite{Jao2005};
(7) \cite{Riedel2010}; 
(8) \cite{Reid1995};
(9) \cite{YPC}}

\end{deluxetable}

  
\subsection{Near-UV Photometry from GALEX}
\label{subsec:GALEX}

We extracted the near-UV (NUV) photometry of SCR 1848 and other red
dwarfs from \cite{Henry2004} and, for comparison, white dwarfs from
\cite{Bergeron2001} from the GALEX General Release 6.  We converted
the apparent magnitudes centered at 2271\AA~to the absolute $M_{2271}$
values given in Table~\ref{tbl:GALEXphot} using distances in
\cite{Bergeron2001}, \cite{YPC} and our parallaxes from CTIOPI.  SCR
1848 is 2.1--3.7 magnitudes brighter than other M4.5V or M5.0V stars
in this band, indicating that at UV wavelengths, the red dwarf is
probably not the primary contributor to the flux, and hence, the one
source seen in both of the HST-STIS UV filters (discussed in
section~\ref{subsec:STISNUV}) is the blue companion, not the red dwarf
(see section~\ref{subsec:STISNUV} for more details).  SCR 1848 also
has a brighter $M_{2271}$ than most of the white dwarfs listed in
Table~\ref{tbl:GALEXphot}.  We note that SCR 1848 is undetected in the
far-UV band at 1528\AA~in GALEX General Release 6.


\begin{table}
\centering
\begin{tabular}{lrccc}
\hline
Object         &   Type/T     & Parallax & $M_V$ &  $M_{2271}$ \\
               &     /K       &  (mas)   & (mag) &  (mag)  \\
\hline
SCR 1848$-$6855&  M5.0V       &   40.6  & 14.9   &  18.9   \\
\hline
GJ~1156        &  M4.5V       &  152.9   & 14.7  &  21.0   \\
GJ~1103        &  M4.5V       &  114.0   & 13.5  &  22.1   \\
GJ~1057        &  M4.5V       &  117.1   & 14.2  &  22.4   \\
LHS~3262       &  M5.0V       &  105.5   & 13.7  &  22.6   \\
Proxima        &  M5.0V       &  774.3   & 15.6  & \nodata \\
\hline
WD~1300$+$263  &      DA/4320 &   28.4   & 16.0  &  16.2   \\
WD~1136$-$286  &  non-DA/4490 &   24.5   & 17.5  &  17.0   \\
WD~0222$+$648  &  non-DA/4520 &   31.4   & 15.8  &  20.0   \\
WD~2251$-$070  &  non-DA/4580 &  123.7   & 16.2  &  23.3   \\
WD~1345$+$238* &      DA/4590 &   82.9   & 15.3  &  21.9   \\
WD~2054$-$050* &  non-DA/4620 &   64.6   & 15.7  &  21.7   \\
WD~2316$-$064  &  non-DA/4720 &   32.2   & 15.7  &  21.3   \\
WD~0029$-$032  &  non-DA/4770 &   42.6   & 15.5  &  19.5   \\
WD~1820$+$609* &  non-DA/4780 &   78.2   & 15.2  &  21.8   \\
WD~2002$-$110  &  non-DA/4800 &   57.7   & 15.8  &  21.1   \\
WD~1444$-$174  &  non-DA/4960 &   69.0   & 15.6  &  21.3   \\
WD~0657$+$320* &      DA/4990 &   53.5   & 15.3  &  17.2   \\
\hline
\end{tabular}

\caption{Absolute V and near-UV magnitudes, $M_{2271}$, from GALEX
  General Release 6 data for our science target SCR 1848, red dwarfs
  with similar spectral types, and white dwarfs cooler than 5000K
  (sorted by temperature).  Asterisks indicate selected cool white
  dwarfs plotted in Figure~\ref{fig:HRUV}.  Red dwarf data are from
  RECONS, while the white dwarf data are from \cite{Bergeron2001}.
  Trigonometric parallaxes are from \cite{Bergeron2001}, the Yale
  Parallax Catalog \citep{YPC}, and our astrometry program.}

\label{tbl:GALEXphot}
\end{table}

\subsection{Red and Blue Spectra from CTIO, Gemini-S, and HST}
\label{sec:redblue}

\begin{figure}
  \centering
  \includegraphics[scale=0.7]{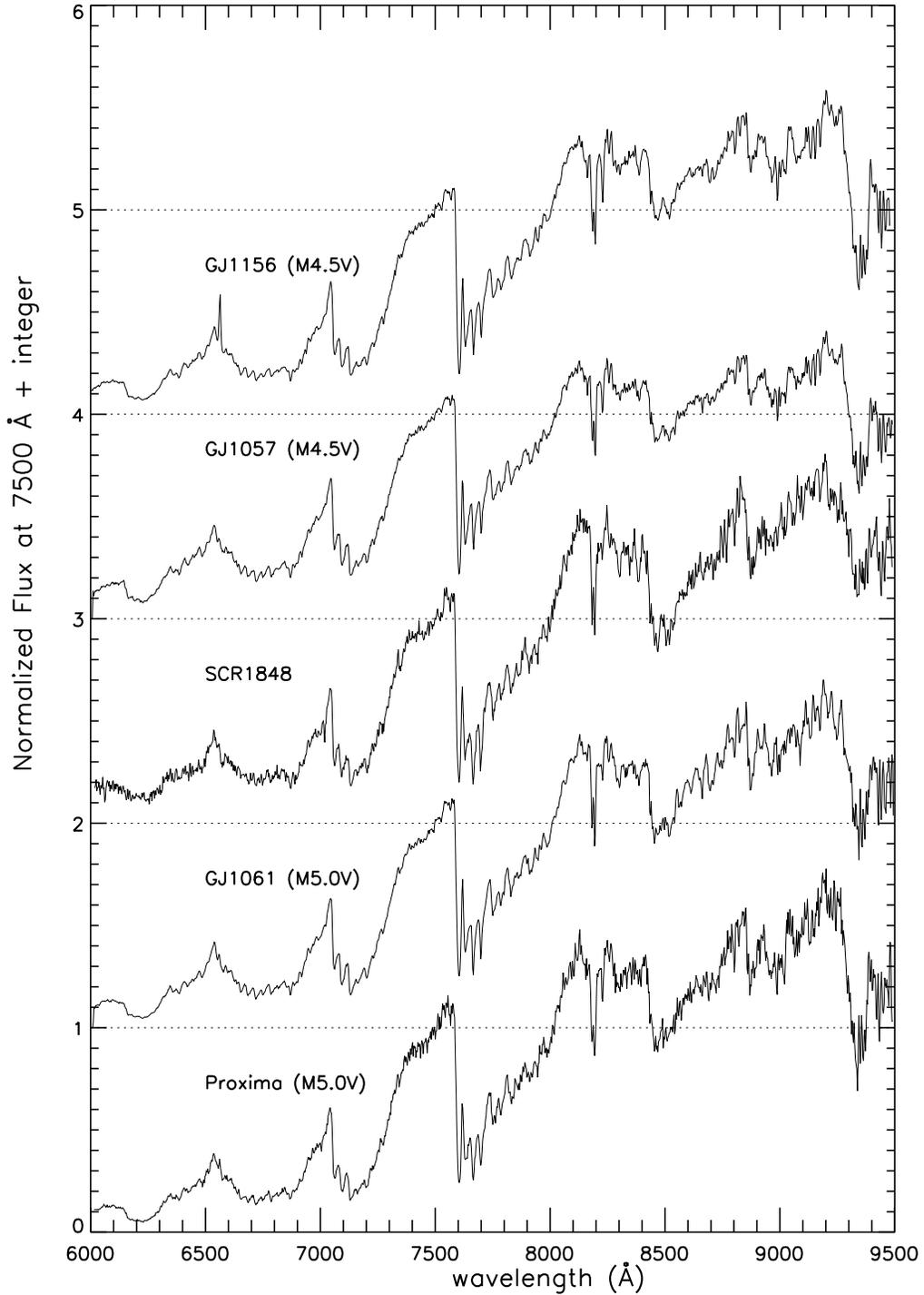}
  
 \caption{The red spectrum of SCR 1848 is shown in the middle of
   spectra of four similar red dwarfs.  Spectra were acquired between
   2003 and 2006 at the CTIO 1.5-m using the R-C Spectrograph, as
   discussed in section~\ref{sec:redblue}.  All spectra are
   normalized at 7500\AA~and offset by unit amounts, where the dotted
   lines are included for guidance.}
 
  \label{fig:sequence}
  \end{figure}

Two red spectra of SCR 1848 covering 6000--9500\AA~were obtained on UT
2003 July 15 using the CTIO 1.5-m telescope and R-C Spectrograph with
the Loral 1200$\times$800 CCD camera.  Observations were made using a
2\arcsec~slit, order-blocking filter OG570, and grating \#32 in first
order with a tilt of 15$\fdg$1.  Bias frames, dome flats, and sky
flats were taken at the beginning of the night for calibration, and
these two exposures of SCR 1848 were taken to permit cosmic ray
rejection.  A 10 second Ne$+$He$+$Ar arc lamp spectrum was recorded
after the science exposures to enable wavelength calibration.  The
spectroscopic flux standard star HR7950 was observed that night.
Reductions were carried out using $IRAF$ reduction packages --- in
particular {\it onedspec.dispcor} for wavelength calibration and {\it
onedspec.calibrate} for flux calibration.  Comparison of this spectrum
with the set of spectral dwarf standards discussed in \cite{Jao2008}
results in a spectral type determination of M5.0V for SCR 1848, with
the usual adopted error of half a sub-type.  The final spectrum is
shown in Figure~\ref{fig:sequence} along with four other red dwarfs of
similar type, all taken at the CTIO 1.5-m using the R-C Spectrograph.
The overall spectral slope and features of SCR 1848 are similar to the
M5.0V standards GJ 1061 and Proxima Centauri at wavelengths between
6900\AA~and 9300\AA.

\begin{figure}
  \centering
  \includegraphics[angle=90, scale=0.7]{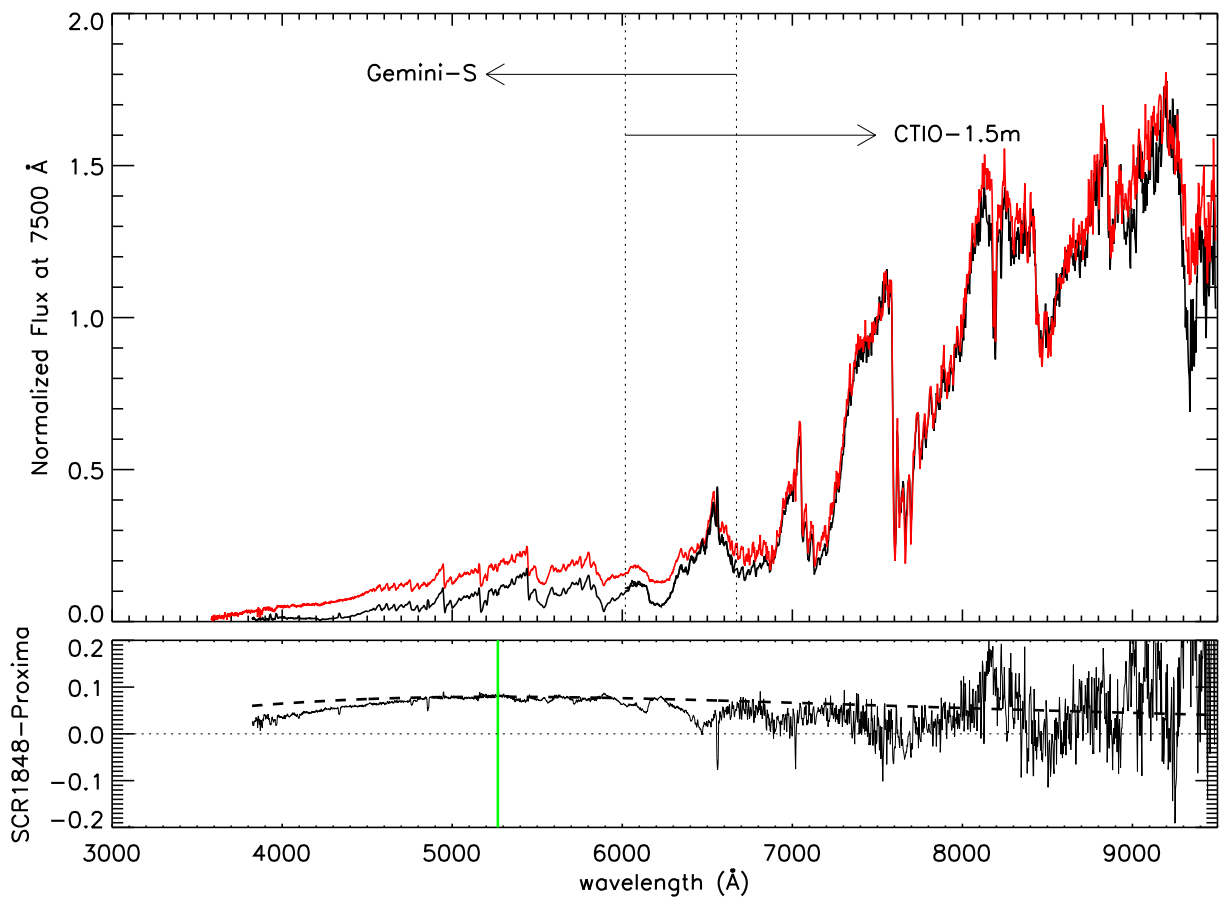}

  \caption{The top panel shows the merged spectra of SCR1848 (red) and
    Proxima (black) from the CTIO 1.5-m and Gemini-South telescopes.
    The spectra are normalized at 7500\AA.  The two vertical dotted
    lines outline the overlapping wavelength regions acquired at both
    telescopes.  The bottom panel shows the flux differences between
    the SCR1848 and Proxima spectra.  The dashed line shows a 5500K
    blackbody curve, and the green line indicates the peak at the
    $\lambda_{max}$ of the difference spectrum.  Balmer absorption
    lines in the difference spectrum are evident because Proxima is
    an active star.}

  \label{fig:SCR1848spectra}
  \end{figure}

Blue spectra of SCR 1848 covering 3580--6670\AA~were obtained on the
night of UT 2010 August 06 using the Gemini-South telescope and GMOS.
Observations were made using a 2\arcsec~slit and grating B600\_G5323.
Two sets of three 600-second exposures were taken during sky
conditions satisfying the 90-percentile image quality criterion, at an
airmass of 1.35.  The two sets of exposures were centered at
5000\AA~and 5250\AA~respectively, so that the CCD gaps on the GMOS
chip could be removed.  Three one-second dome flats were taken
immediately after each set of exposures and used for flat-fielding.
Wavelength calibration exposures of the CuAr arc lamp were part of
Gemini baseline calibrations and were taken at the end of the night.
A spectroscopic standard star, LTT7987, was observed with the exact
same setups before observing SCR 1848.  The final blue spectrum of SCR
1848 combined from all calibrated nodding spectra is shown in
Figure~\ref{fig:SCR1848spectra}. The long-slit spectra were reduced
using instructions presented during the Gemini Data Reduction Workshop
({\it www.noao.edu/meetings/gdw/}).

Using an overlap region spanning $\sim$700\AA, the red and blue
spectra of SCR 1848 were merged and are shown with the red line in
Figure~\ref{fig:SCR1848spectra}.  The black line is Proxima, an M5.0V
spectral standard star.  Proxima's red spectra were obtained on UT
June 6 2006 at the CTIO 1.5-m using the same R-C Spectrograph setup
and procedures as for SCR 1848, whereas the blue spectra were obtained
on UT 2011 May 7 at Gemini-South using the same GMOS setup discussed
above, but under poor weather conditions.  Because Proxima is very
bright at optical wavelengths ($V =$ 11.13), the integration times
were only 30 seconds on Gemini-South.  The flux differences between
SCR 1848 and Proxima (SCR1848$-$Proxima$_{ground}$) are shown in the
bottom panel of Figure~\ref{fig:SCR1848spectra}.

Spectroscopic observations of Proxima from HST/MAST were extracted to
provide an additional comparison to our SCR 1848 spectra and to
double-check our ground-based Proxima spectra.  A blue spectral
observation of Proxima (Y2WY0305T) was made in 1996 with HST's Faint
Object Camera (FOC) with an exposure time of 430 seconds in the G570H
filter, with wavelength coverage of 4569--6817\AA.  A red spectral
observation of Proxima (Y2WY0705T) was also made by the FOC in 1996,
with an exposure time of 280 seconds in the G780H filter, with
wavelength coverage of 6269--8498\AA.  As before, we use an overlap
region spanning $\sim$500\AA~to merge the red and blue spectra from
HST.

\begin{figure}
  \centering
  \includegraphics[angle=90, scale=0.7]{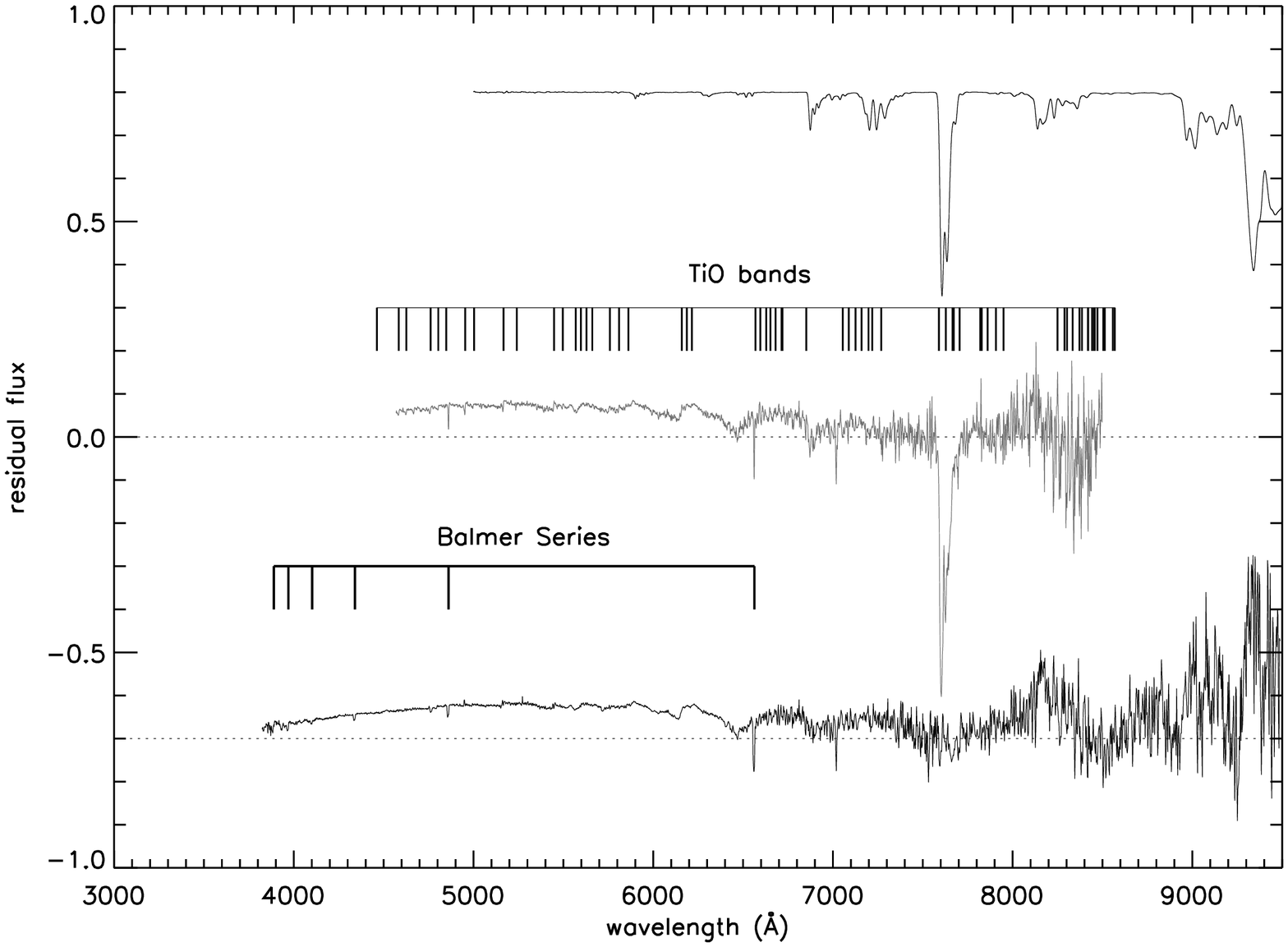}

  \caption{The top thin line shows telluric absorptions from
    \cite{Hinkle2003}.  Various TiO band absorption features from
    \cite{Turnshek1985} are shown as black tick marks.  The gray and
    black lines indicate the excess flux of SCR 1848 compared to
    Proxima based on the HST (SCR1848$-$Proxima$_{HST}$) and
    ground-based (SCR1848$-$Proxima$_{ground}$) observations,
    respectively.  Note the blue excess in SCR1848 are in both
    subtractions, which are virtually identical.  Proxima has strong
    Balmer emission lines (see the guide plotted), which results in
    absorptions in the subtracted spectra.}
  
  \label{fig:HSTGMOS}
  \end{figure}

Shown in Figure~\ref{fig:HSTGMOS} are the flux differences between SCR
1848 and Proxima using both the HST spectra for Proxima (gray line,
SCR 1848$-$Proxima$_{HST}$) and our ground-based spectra (black line,
SCR 1848$-$Proxima$_{ground}$, same as in the bottom panel of
Figure~\ref{fig:SCR1848spectra}).  The flux differences are virtually
identical, other than in the telluric bands outlined in the spectrum
at the top of the plot.  Note that the Balmer lines appear as
absorption features in the differential spectra because in Proxima
those lines are in emission.  These matching results give us
confidence that our ground-based observations of Proxima, and
presumably SCR 1848 as well given that it was observed using the same
instrumental setup, are reliable.  Mismatches redward of 6000\AA~are
due primarily to differences in the precise strengths of the TiO bands
between Proxima and SCR 1848.  We see similar mismatches when
comparing two M5.5V dwarfs.  Blueward of $\sim$6500\AA, the residual
spectrum is evidently due to the unseen companion, and is effectively
featureless.

  
\clearpage

\subsection{Resolution of the Binary and Near-UV Photometry from HST}
\label{subsec:STISNUV}

\subsubsection{Matching sources on two HST-STIS images}

\begin{figure}
  \centering
  \includegraphics[angle=90, scale=0.7]{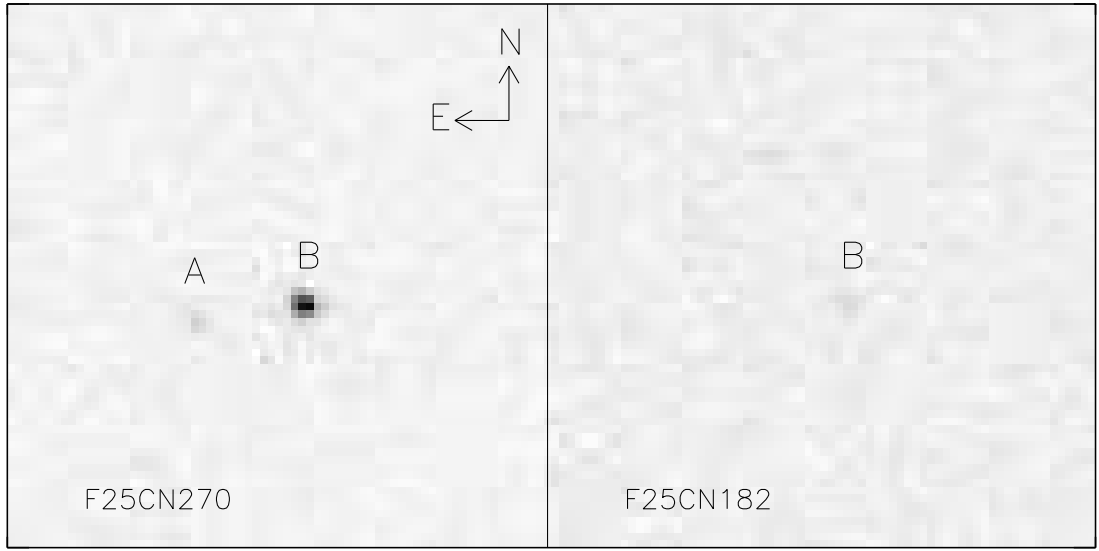}
  
  \caption{The left panel shows two sources in the field taken
  through filter F25CN270 using HST-STIS$+$MAMA.  The right panel
  shows the sole source in the F25CN182 filter.  The details of
  assigning components A and B to the sources are discussed in
  section~\ref{subsec:STISNUV}.  Both fields are 1\farcs7 on a
  side.}
  
  \label{fig:STIS.MAMA}
  \end{figure}

In an attempt to resolve the system, we imaged SCR 1848 on UT 2011
July 14 using HST-STIS$+$MAMA in two UV filters --- F25CN270 at
2711\AA~and F25CN182 at 2004\AA.  The field of view was
25\arcsec$\times$25\arcsec~with a pixel scale of 24.6 mas/pixel. The
observation logs, derived astrometry, and resulting photometry are
given in Table~\ref{tbl:STIS.coords.mags}.  Figure~\ref{fig:STIS.MAMA}
shows cropped images 1\farcs7 on a side from HST-STIS$+$MAMA,
revealing two sources in the F25CN270 filter, but only one very faint
source in F25CN182.  We note that there were no other sources detected
in the full field in either filter.

Given the lack of other stars for reference in the field, we need to
determine carefully which source is the M5.0V star and which is the
previously unseen companion.  First, we used the {\it xy2rd} task from
the HST/STSDAS IRAF package to determine the coordinates of both
sources in the F25CN270 image, yielding the results given in
Table~\ref{tbl:STIS.coords.mags}.  The coordinates for the sources in
the two F25CN270 images differ by $\sim$0.001 seconds in RA and
0\farcs004--0\farcs009 in DEC, where the differences come from a
combination of the HST pointing accuracy, the image distortion
correction and SExtractor's centroiding accuracy.  Because the
separation between the two sources is 0\farcs349, which is much larger
than the errors in the positions, and because the sole source in the
F25CN182 image has a similar position as the bright source in the
F25CN270 image, we conclude that the bright source is in both F25CN270
and F25CN182 images.

\subsubsection{Identifying the red dwarf on HST-STIS images}

To identify which source corresponds to which component, we explore
what an M5.0 dwarf's near-UV brightness is likely to be in the
HST-STIS images.  Proxima Centauri is an M5.0 dwarf similar to SCR
1848, and has IUE archived spectra covering 1851--3347\AA~that were
extracted from the IUE/MAST archives.  There are a total of 41 spectra
of Proxima taken between 1979 and 1995.  In this wavelength range,
Proxima has no detected continuum flux, and the only prominent feature
-- the Mg II line at 2798\AA~-- is stable throughout these 41 spectra.
\cite{Buccino2008} indicate that both Mg II and Ca II H$+$ K lines are
good indicators of the thermal structures of stellar atmospheres and
\cite{Cincunegui2007} demonstrated a linear relation between Ca II and
H$\alpha$. This implies a likely relation between Mg II and H$\alpha$
lines. At optical wavelengths, Proxima exhibits a strong H$\alpha$
emission feature and shows long-term variability in the $V$ band at a
level of 29 milli-mags over 10.9 years\footnote{The variability of red
dwarfs and cool subdwarfs using our astrometry frames is discussed in
\cite{Jao2011}.}. This indicates that Proxima is active.

In contrast, we do not see strong H$\alpha$ emission from SCR 1848 in
its red spectrum so we expect SCR 1848 would have a weaker Mg II line
than Proxima. The long-term variability of SCR 1848 in the $I$ band is
only 9 millmags over 8.5 years, although this lower level might be
expected because red dwarfs generally vary more in the $V$ band than
in the $I$ band.  However, four nights of $VRI$ photometry, which
includes both internal (signal-to noise, variability) and external
(extinction corrections, photometric standard star stability) errors
for SCR 1848 show a variability of only 15 milli-mags in the $V$ band,
only half that seen in Proxima's long-term measurements.  We conclude
that SCR 1848 (a) has a quieter atmosphere than Proxima, (b) has
weaker Mg II emission, and (c) should have less flux than Proxima from
1851--3347\AA.  Finally, as can be seen in
Figure~\ref{fig:SCR1848spectra}, the companion's flux is significant
blueward of $\sim$6500\AA, and we anticipate that at the HST-STIS
wavelengths, the blue component dominates the waning flux of the red
dwarf.  We therefore conclude that the source seen in both HST-STIS
bands is the blue companion, and the fainter source seen in only the
F25CN270 image is the red dwarf.  The mean separation and position
angle of companion measured from two F25CN270 images are 0\farcs349
and 277$\fdg$6 relative to the M5.0 dwarf at the mean epoch of
2011.5332. Also, it shows that faint background star discussed in
section~\ref{sec:background} observed in 2011 is not the same as this
component because the background star is beyond the field of view of
STIS$+$MAMA.

\subsubsection{NUV Photometry from HST-STIS images}

We performed aperture photometry on the HST-STIS images using
SExtractor \citep{sextractor}.  To determine photometry in the STMAG
system, we calculated $f_{\lambda}$ based on the detected counts and
image exposure times, and used the equation ($STMAG=-2.5 \log
f_{\lambda}-21.1$) outlined in \S5.3 of the STIS Data Handbook
\citep{Bostroem2011}.  Because the two sources are separated by only
14 pixels, we selected a standard 3-pixel radius aperture for the
photometry and used the correction in Table 8 of the Instrument
Science Report STIS 2003-001 \citep{Proffitt2003} to derive
aperture-corrected photometry.  The mean $m_{2711}$ values from the
two F25CN270 images for A and B are 23.60$\pm$0.17 and 20.91$\pm$0.03,
respectively, where A is the red dwarf.  The mean $m_{2004}$ for
component B, the previously unseen companion and the sole object in
the field, is 22.79$\pm$0.26.

\begin{deluxetable}{ccclcccccc}
\rotate
\tablewidth{0pt}
\tabletypesize{\footnotesize}
\tablecaption{Results from STIS images \label{tbl:STIS.coords.mags}}
\tablehead{
\colhead{Image}       &
\colhead{UT}          &
\colhead{exposure}    &       
\colhead{source}      & 
\colhead{X}           &
\colhead{Y}           &
\colhead{separations} &
\colhead{RA}          &
\colhead{Decl.}         &
\colhead{mag}         \\
\colhead{}            &
\colhead{}            &  
\colhead{(seconds)}     &
\colhead{}            &
\multicolumn{2}{c}{(pixel)} &
\colhead{(mas)}            & 
\colhead{}            & 
\colhead{}            &  
\colhead{(mean)}     \\
}
\startdata
\multicolumn{9}{c}{F25CN270} \\
    1           & 2011/07/14/20:26:00       &  3086         &  B           & 586.648  &  566.550 & 267/217      & 18:48:20.1958 & $-$68:55:48.705 &  20.91$\pm$0.03 \\
                &                           &               &  A           & 575.812  &  557.713 &              & 18:48:20.2593 & $-$68:55:48.754 &  23.60$\pm$0.17 \\
\tableline
    2           & 2011/07/14/21:18:14       &  3349         &  B           & 596.849  &  566.372 & 278/219      & 18:48:20.1948 & $-$68:55:48.714 &        \\
                &                           &               &  A           & 585.552  &  557.481 &              & 18:48:20.2602 & $-$68:55:48.758 &        \\
\tableline
\tableline
\multicolumn{9}{c}{F25CN182} \\
     1          & 2011/07/14/22:15:10       &  3330         &  B           & 588.237  &  567.504 &              & 18:48:20.1873 & $-$68:55:48.705 &  22.79$\pm$0.26 \\
\tableline
     2          & 2011/07/14/23:11:28       &  3349         &  B           & 597.885  &  567.686 &              & 18:48:20.1876 & $-$68:55:48.699 &        \\
\enddata

\tablecomments{Pixel coordinates and RA and DEC for objects detected
  in the HST-STIS$+$MAMA images are given, as measured using
  SExtractor and the STSDAS {\it xy2rd} task.  The separations in mas
  are given in the X and Y directions.  The mean STMAG magnitudes of
  two sources in F25CN270 and the only source in F25CN182 are listed
  in the last column. The components are labelled as A and B in
  Figure~\ref{fig:STIS.MAMA}.}

\end{deluxetable}

  
\section{Discussion}

\subsection{The Companion's Temperature and Spectral Type}

From the various observations of SCR 1848 outlined above, we have (1)
detected a large perturbation on an M5.0V star at 24.6 pc caused by an
unseen companion, (2) detected the excess $U$ band and NUV flux from
the companion, (3) measured excess flux spectroscopically blueward of
$\sim$6500\AA, and (4) resolved the binary system.  The next step is
to determine the nature of the companion.

\begin{figure}
  \centering
  \includegraphics[angle=90,scale=0.7]{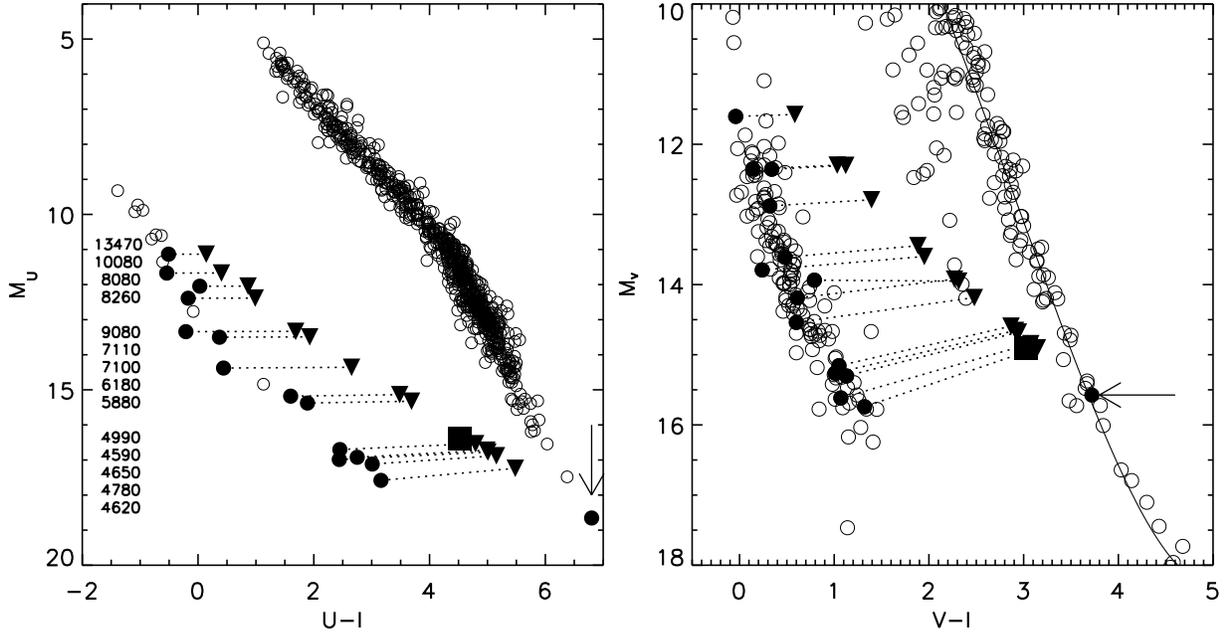}
  \vspace*{-2 cm}
  \caption{Photometry of simulated binaries composed of the M5.0V red
  dwarf Proxima (single, indicated by arrow) and various white dwarfs
  (other filled circles).  The location of the SCR 1848 system
  (combined light) is indicated by a filled box.  The filled triangles
  indicate the combined photometry of Proxima and each white dwarf
  ``companion'', with dashed lines connecting the components of each
  simulated system.  Effective temperatures are listed to the left of
  the corresponding white dwarfs ordered by their absolute magnitudes.
  Open circles are main sequence dwarfs, subdwarfs, and white dwarfs
  from \cite{Koen2010}, \cite{Bergeron2001} and RECONS, e.g.,
  \cite{Jao2005} and \cite{Henry2006}.  The fit line in the right
  panel represents the stellar main sequence.}
  \label{fig:HRUV}
  \end{figure}

We simulated $UVI$ photometry for systems composed of the M5.0V red
dwarf Proxima and various white dwarfs with $UVI$ from
\cite{Bergeron2001}.  Basically, we move Proxima and white dwarfs to
SCR 1848's distance at $\sim$24.6 pc but keep the same absolute
magnitudes. We calculate their combined flux and photometry in
different filters to simulate unresolved M5.0V$+$WD binaries. The
results are shown in Figure~\ref{fig:HRUV}.  We found that to
reproduce the combined $UVI$ photometry for SCR 1848, a white dwarf
companion would need to have T$_{\rm eff}$$\sim$5000K to have enough
flux for the combined U photometry, but T$_{\rm eff}$$\sim$4600K to
match the flux indicated by the combined $V$ photometry.  These
results indicate that the blue companion could be a white dwarf with
T$_{\rm eff}$$\sim$4600--5000K\footnote{There are only five white
dwarfs cooler than 5000K in \cite{Bergeron2001} having $U$ band
photometry, which is from the White Dwarf Catalog
\citep{McCook1999}.}.

  
Given the resolution of the system via HST-STIS, we can estimate the
temperature of the companion by generating a grid of model NUV colors
($m_{F25CN182}-m_{F25CN270}$) under the assumption that the secondary
emits as a blackbody.  To find the colors (observed$-$modeled) for
given $V$ magnitudes, we used the {\it calcphot} task in the IRAF
STSDAS/SYNPHOT package to generate synthetic photometry for the
secondary.  Based on the results shown in Figure~\ref{fig:HRUV}, we
explore temperatures of 4000K--5000K, and target $V$ magnitudes of
16--19 that correspond to the brightness anticipated at the distance
of the SCR 1848 system.  Our results indicate that the best match
between the observed and modeled NUV colors
($\delta$NUV$_{model-observed}=$0.03 mag) are for a blackbody with
$T=4500$K, $V=18.4$, and $U=19.4$ at the distance of SCR
1848. However, a known DA white dwarf with $T=4590$K, WD 1345$+$238,
would have $V=$17.2 and $U=$18.8, if it were moved to the distance of
SCR 1848. It appears that the modeled white dwarf $U$ and $V$
magnitudes for SCR 1848 B are fainter than a typical field white dwarf
with $T$$\sim$4600K. To complicate matters further, as shown in the
bottom panel of Figure~\ref{fig:SCR1848spectra}, the residual flux
peaks at $\sim$5300\AA, corresponding to the peak of a 5500K blackbody
curve, nearly 1000K hotter.  So, the exact temperature and nature of
the companion remains elusive and a blackbody curve may not be ideal
to represent its spectrum, but its temperature is likely 4600--5500K.

The bottom panel of Figure~\ref{fig:SCR1848spectra} also shows that
the companion's spectrum appears to be featureless.  According to
\cite{McCook1999}, a DC white dwarf is defined to have a continuous
spectrum with no lines deeper than 5\% in any part of the
electromagnetic spectrum.  This definition appears to be consistent
with the spectroscopic information we have for the companion.  It
implies a spectral type of DC if it is in fact a white dwarf. In the
inferred temperature range (i.e., T$_{\rm eff}$ $<$ 5500 K), the white
dwarf will show different continuum features in the optical depending
on atmospheric composition. The possible atmospheric compositions are
1) pure hydrogen, where collision induced absorption (CIA) occurs, 2)
pure helium, where CIA is unimportant, or 3) mixed hydrogen and
helium, where CIA occurs at hotter temperatures because of collisions
with neutral helium \citep{Bergeron2002}. Without knowing which model
to apply, and how much the CIA affects its spectral energy
distribution, the effective temperature remains uncertain, but is
likely in the temperature range discussed above. Good S/N resolved
photometry covering most of the SED of the companion is needed to best
constrain the effective temperature \citep{Kilic2010}. For now, we
estimate that the companion may be a DC type white dwarf.

\subsection{The Companion's Mass}
\label{sec:companion}

The perturbation curve shown in Figure~\ref{fig:1a} maps the shift in
position of the binary system's photocenter and reveals information
about the masses of the components.  However, because our data do not
yet cover a full orbital period for the system, we find that the
perturbation curve is not stable, with a slope in the curve after 2006
that varies as more images are acquired.  An additional {\it caveat}
is available from the HST-STIS image, in which the B component is
observed at a position angle of 277$\fdg$6 in mid-2011, where the
position angle of the photocenter from our perturbation curve lies at
63$\fdg$3.  The position angle of the secondary should be
180$^{\circ}$ from the position of the photocenter offset,
corresponding to 243$\fdg$3 rather than 277$\fdg$6, or a 34$\fdg$3
difference.  We could shift residuals or the photocenter to match the
position angle from STIS images, but we would not be significantly
more confident in the orbit. We expect this mismatch is because of the
incomplete photocentric orbit from our astrometric observations.


\begin{deluxetable}{ccccccc}
\tablewidth{0pt}
\tablecaption{The preliminary orbital elements of SCR1848's photocenter\label{tbl:element}}
\tablehead{
\colhead{P}           &    
\colhead{$\alpha$}         &  
\colhead{$i$}         &
\colhead{$e$}         &    
\colhead{T$_{0}$}     &  
\colhead{$\omega$}    &
\colhead{$\Omega$}   \\
\colhead{(yr)}        & 
\colhead{($\arcsec$)} & 
\colhead{($^{\circ}$)}  &          
\colhead{}            &   
\colhead{(epoch)}     &  
\colhead{($\circ$)}   &  
\colhead{($\circ$)}
}
\startdata
15.17       &  0.2082   &  90.32   &  0.84    &    2004.58   &  280.74     &   248.08   \\
$\pm$3.25   &  $\pm$0.0255   &  $\pm$0.37    &  $\pm$0.03    &    $\pm$0.02      & $\pm$4.85       &   $\pm$2.25
\enddata
\end{deluxetable}

\begin{figure}
  \centering
  \includegraphics[scale=0.7]{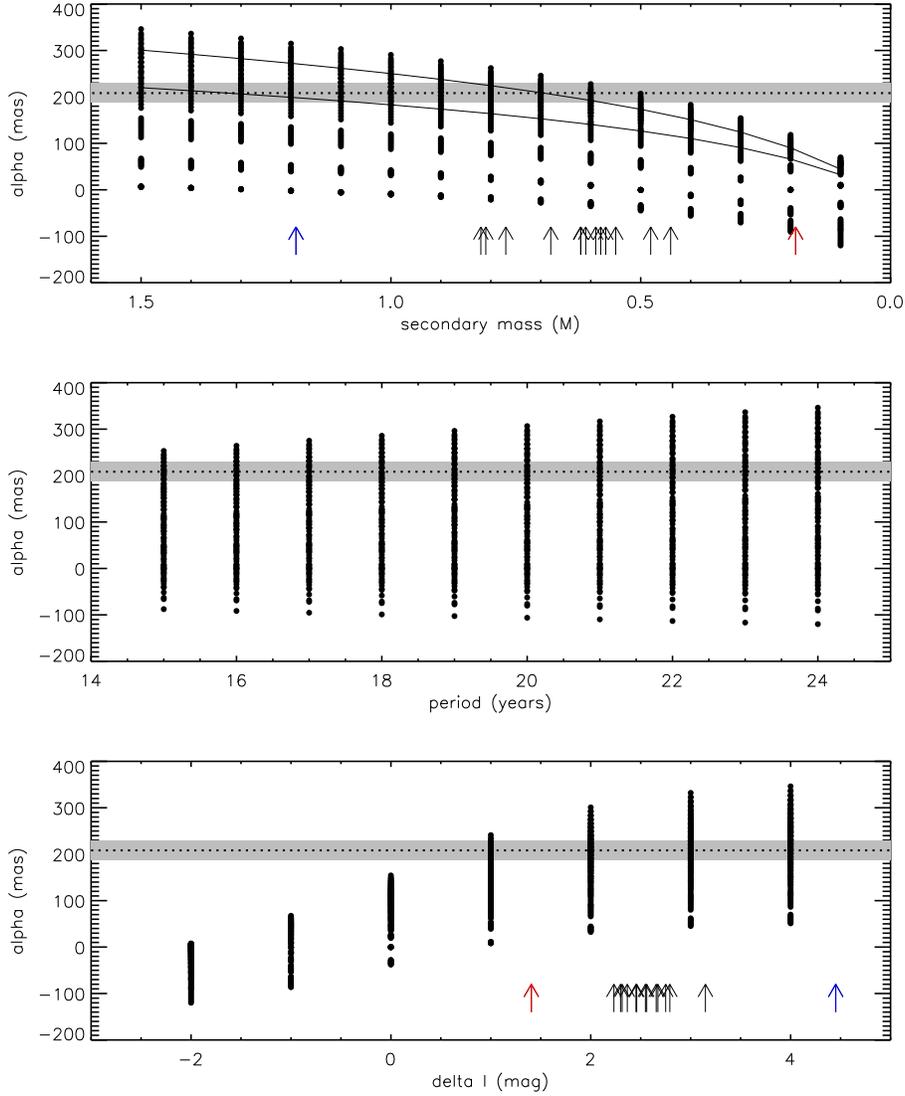}

  \caption{The modeled semi-major axes, $\alpha_{model}$, of the
  photocentric orbit, are plotted against various secondary masses,
  periods, and $\Delta$$I$ values.  Dotted lines at 208 mas represent
  SCR 1848's $\alpha_{fit}$, determined from fitting the parallax
  residuals, while the gray boxes represent $\pm$10\% of
  $\alpha_{fit}$. This region outlines what is likely to be the
  minimum $\alpha_{fit}$.  Dots are simulated $\alpha_{model}$ from
  various cases discussed in the text.  In the top panel, the top and
  bottom solid lines indicate binaries with a fixed $\Delta$$I=$2
  having $P=$15 (top line) and 24 (bottom line) years, respectively.
  The arrows indicate 15 white dwarfs with $T =$ 4490--5000K from
  \cite{Bergeron2001}.  The red arrow represents WD 1136$-$286, and
  the blue arrow represents WD 0324$+$738.  Both are extreme cases of
  cool white dwarfs. }

  \label{fig:model.alpha}
  \end{figure}

With these reservations in mind, we fit a photocentric orbit given in
Table~\ref{tbl:element} using the methodology outlined in
\cite{Hartkopf1989}.  Given the derived inclination is close to
90$^{\circ}$ or edge-on, these orbital elements are suspect. However,
one thing is certain --- the orbital period is longer than our
observing sequence of 8.5 years, and is otherwise poorly constrained.
From the preliminary fit to the data, we find that the semi-major axis
of the photocentric orbit is at least $\alpha_{fit}$ $=$ 208 mas.  We
consider this a minimum semi-major axis given that the perturbation
amplitude continues to grow as more frames are taken.  We can use this
value to constrain the mass of the companion by modeling
$\alpha_{fit}$ using $\alpha=(B-\beta)a$, where $B$ is the fractional
mass ($M_{B}/(M_{A}+M_{B})$), $\beta$ is the relative luminosity
($1/(1+10^{(0.4)\Delta m})$), and $a$ is the semi-major axis of the
relative orbit of the two components \citep{vandekamp}.  $\Delta m$ is
the magnitude difference in the $I$ band in this case, because we used
the $I$ filter for the astrometric series.  The modeled results are
shown in Figure~\ref{fig:model.alpha}, in which possible secondaries
have masses 0.1--1.5$M_{\odot}$, are fainter by 0--4 mag at $I$, and
are in orbits with periods of 15--24 years.  When constrained by
$\alpha_{fit}$ $\ge$ 208 mas from our preliminary fit, we see that the
secondary has $M \ge 0.5M_{\odot}$ and $\Delta I \ge $ 1.  This
implies that the secondary is too massive to be brown dwarf, which is
also consistent with our discovery that the secondary is bluer than
the M5.0V primary at optical/UV wavelengths.  The period, as expected,
is poorly constrained, and of course limits our ability to measure the
companion mass.  For example, if the secondary has a period of 15
years and $\Delta I=$2 (top solid line in the top panel of
Figure~\ref{fig:model.alpha}), its mass is $\sim$0.7$M_{\odot}$. If
the secondary has a period of 24 years and $\Delta I=$2 (bottom solid
line in top panel of Figure~\ref{fig:model.alpha}), its mass is
$\sim$1.3$M_{\odot}$.  As a final consideration of the companion's
mass, we compare 15 known cool white dwarfs with $T =$ 4490--5000 K
from \cite{Bergeron2001}.  Figure~\ref{fig:model.alpha} shows most of
these cool white dwarfs have masses of 0.4--0.9$M_{\odot}$, and that
their I magnitudes would be 2--3 magnitudes fainter than SCR 1848
after being placed at SCR 1848's distance.  Thus, we cannot constrain
the mass at this time, so it is important that we continue to follow
the system to determine an accurate mass for the companion, which
would provide a rare mass measurement of a low-temperature white
dwarf.

  
\section{Conclusions}

We present results of a nearby star system, SCR 1848, which has the
largest perturbation among more than 700 targets observed in our
long-term astrometry program at the CTIO 0.9-m.  We present a suite of
data from the CTIO 0.9-m, CTIO 1.5-m, Gemini-South Observatory, the
NASA Galaxy Evolution Explorer, the International Ultraviolet
Explorer, and Hubble Space Telescope, to study the nature of this
intriguing system.  Results show that this M5.0 dwarf has a blue
excess that probably comes from a faint white dwarf with an effective
temperature of 4600--5500K and a mass greater than 0.5$M_{\odot}$.  We
have resolved the system using HST-STIS$+$MAMA, which provides hope
that direct spectroscopic and photometric observations of the
secondary are possible, and thereby a determination of its spectral
type. Such low temperature and intrinsically faint white dwarfs are
rarer than their hotter counterparts. For example, \cite{Rebassa2010}
reported 1602 white-dwarf plus main-sequence star binaries from the
Sloan Digital Sky Survey Data Release 6, but none of those white
dwarfs nor main sequence stars is cooler than 6000K (see their Table
7). This work provides an alternative method for detecting low
temperature white dwarfs. Given the large astrometric perturbation
this type of system produces, it would be easy for astrometric
programs like Pan-STARRS, Gaia or Large Synoptic Survey Telescope to
detect such systems. Despite more than 8 years of astrometric
observations, this system has not yet completed a full orbit, which
limits our ability to determine accurate photocentric orbital elements
and determine the component masses.  We intend to continue observing
SCR1848 until future efforts can provide crucial dynamical masses of a
low temperature white dwarf and a low mass red dwarf.

\section{Acknowledgments}

The astrometric observations reported here began as part of the NOAO
Surveys Program in 1999 and continued on the CTIO 0.9-m via the SMARTS
Consortium starting in 2003.  We gratefully acknowledge support from
the National Science Foundation (grants AST 05-07711 and AST
09-08402), and NASA's Space Interferometry Mission, which together
have made this long-term effort possible.  We also thank the members
of the SMARTS Consortium, who enable the operations of the small
telescopes at CTIO, as well as the members of the observing support
team, specifically Edgardo Cosgrove, Arturo G\'{o}mez, Alberto
Miranda, and Joselino V\'{a}squez.  We also thank Valery Suleimanov
for discussions regarding the excess flux.

The NUV photometry is based on observations made with the NASA Galaxy
Evolution Explorer.  GALEX is operated for NASA by the California
Institute of Technology under NASA contract NAS5-98034.

The Gemini observations were accomplished via the Gemini Observatory
(GS-2010B-Q-15 and GS-2011A-Q-92), which is operated by the
Association of Universities for Research in Astronomy, Inc., under a
cooperative agreement with the NSF on behalf of the Gemini
partnership: the National Science Foundation (United States), the
National Research Council (Canada), CONICYT (Chile), the Australian
Research Council (Australia), Minist\'{e}rio da Ci\^{e}ncia,
Tecnologia e Inova\c{c}\~{a}o (Brazil) and Ministerio de Ciencia,
Tecnolog\'{i}a e Innovaci\'{o}n Productiva (Argentina).

The IUE data and archival spectra of Proxima presented in this paper
were obtained from the Multimission Archive at the Space Telescope
Science Institute (MAST).  STScI is operated by the Association of
Universities for Research in Astronomy, Inc., under NASA contract
NAS5-26555.  Support for MAST for non-HST data is provided by the NASA
Office of Space Science via grant NAG5-7584 and by other grants and
contracts.

The HST-STIS observations were supported for program number 12259 by
NASA through a grant from the Space Telescope Science Institute, which
is operated by the Association of Universities for Research in
Astronomy, Inc., under NASA contract NAS5-26555.

This research has made use of the SIMBAD database, operated at CDS,
Strasbourg, France.  This work also has used data products from the
Two Micron All Sky Survey, which is a joint project of the University
of Massachusetts and the Infrared Processing and Analysis Center at
California Institute of Technology funded by NASA and NSF.  




\begin{thebibliography}{}

\bibitem[Bergeron et al.(2001)]{Bergeron2001} Bergeron, P., Leggett,
  S.~K., \& Ruiz, M.~T.\ 2001, \apjs, 133, 413

\bibitem[Bergeron \& Leggett(2002)]{Bergeron2002} Bergeron, P., \&
Leggett, S.~K.\ 2002, \apj, 580, 1070



\bibitem[Bertin \& Arnouts(1996)]{sextractor} Bertin, E.~\& Arnouts,
  S.\ 1996, \aaps, 117, 393


\bibitem[Bostroem \& Proffitt(2011)]{Bostroem2011} Bostroem, K.  \&
  Proffitt, C., STIS Data Handbook, Version 6.0, 2011,
  (Baltimore:STScI)

\bibitem[Buccino \& Mauas(2008)]{Buccino2008} Buccino, A.~P.,
\& Mauas, P.~J.~D.\ 2008, \aap, 483, 903



\bibitem[Cincunegui et al.(2007)]{Cincunegui2007} Cincunegui, C.,
D{\'{\i}}az, R.~F., \& Mauas, P.~J.~D.\ 2007, \aap, 461, 1107


\bibitem[Dieterich et al.(2012)]{Deterich2012} Dieterich, S.~B., 
Henry, T.~J., Golimowski, D.~A., Krist, J.~E., 
\& Tanner, A.~M.\ 2012, \aj, 144, 64 


\bibitem[Hambly et al.(2004)]{Hambly2004} Hambly, N.~C., Henry, T.~J.,
  Subasavage, J.~P., Brown, M.~A., \& Jao, W.-C.\ 2004, \aj, 128, 437

\bibitem[Hartkopf et al.(1989)]{Hartkopf1989} Hartkopf, W.~I.,
  McAlister, H.~A., \& Franz, O.~G.\ 1989, \aj, 98, 1014


\bibitem[Hawley et al.(1996)]{Hawley1996} Hawley, S.~L., Gizis, J.~E.,
  \& Reid, I.~N.\ 1996, \aj, 112, 2799

\bibitem[Henry et al.(2004)]{Henry2004} Henry, T.~J., Subasavage,
  J.~P., Brown, M.~A., et al.\ 2004, \aj, 128, 2460

\bibitem[Henry et al.(2006)]{Henry2006} Henry, T.~J., Jao, W.-C.,
  Subasavage, J.~P., Beaulieu, T.~D., Ianna, P.~A., Costa, E., \&
  M{\'e}ndez, R.~A.\ 2006, \aj, 132, 2360

\bibitem[Hinkle et al.(2003)]{Hinkle2003} Hinkle, K.~H., Wallace, L.,
  \& Livingston, W.\ 2003, Bulletin of the American Astronomical
  Society, 35, 1260

\bibitem[Jao et al.(2005)]{Jao2005} Jao, W.-C., Henry, T.~J.,
  Subasavage, J.~P., Brown, M.~A., Ianna, P.~A., Bartlett, J.~L.,
  Costa, E., \& M{\'e}ndez, R.~A.\ 2005, \aj, 129, 1954

\bibitem[Jao et al.(2008)]{Jao2008} Jao, W.-C., Henry, T.~J.,
  Beaulieu, T.~D., \& Subasavage, J.~P.\ 2008, \aj, 136, 840

\bibitem[Jao et al.(2011)]{Jao2011} Jao, W.-C., Henry, T.~J.,
  Subasavage, J.~P., et al.\ 2011, \aj, 141, 117

\bibitem[Kilic et al.(2010)]{Kilic2010} Kilic, M., Leggett, S.~K.,
Tremblay, P.-E., et al.\ 2010, \apjs, 190, 77



\bibitem[Koen et al.(2010)]{Koen2010} Koen, C., Kilkenny, D., van Wyk,
  F., \& Marang, F.\ 2010, \mnras, 403, 1949

\bibitem[McCook \& Sion(1999)]{McCook1999} McCook, G.~P., \& Sion,
  E.~M.\ 1999, VizieR Online Data Catalog, 3210, 0

\bibitem[Proffitt et al.(2003)]{Proffitt2003} Proffitt, C., Brown,
  T.~M., Mobasher, B., and Davies J., 2003, Instrumental Science
  Report, 2003-001, (Baltimore:STScI)

\bibitem[Rebassa-Mansergas et al.(2010)]{Rebassa2010}
Rebassa-Mansergas, A., G{\"a}nsicke, B.~T., Schreiber, M.~R., Koester,
D., \& Rodr{\'{\i}}guez-Gil, P.\ 2010, \mnras, 402, 620

\bibitem[Reid et al.(1995)]{Reid1995} Reid, I.~N., Hawley, 
S.~L., \& Gizis, J.~E.\ 1995, \aj, 110, 1838 


\bibitem[Riedel et al.(2010)]{Riedel2010} Riedel, A.~R., Subasavage,
  J.~P., Finch, C.~T. et al. \ 2010, \aj, 140, 897

\bibitem[Riedel et al.(2011)]{Riedel2011} Riedel, A.~R., Murphy,
  S.~J., Henry, T.~J., et al.\ 2011, \aj, 142, 104

\bibitem[Skrutskie et al.(2006)]{2mass} Skrutskie, M.~F., Cutri,
R.~M., Stiening, R. et al.\ 2006, \aj, 131, 1163


\bibitem[Subasavage et al.(2009)]{Subasavage2009} Subasavage, J.~P.,
  Jao, W.-C., Henry, T.~J., Bergeron, P., Dufour, P., Ianna, P.~A.,
  Costa, E., \& M{\'e}ndez, R.~A.\ 2009, \aj, 137, 4547

\bibitem[Turnshek et al.(1985)]{Turnshek1985} Turnshek, D.~E.,
  Turnshek, D.~A., \& Craine, E.~R.\ 1985, An Atlas of Digital Spectra
  of Cool Stars, (Tucson: Western Research Company)

\bibitem[van Altena et al.(1995)]{YPC} van Altena, W.~F., Lee, J.~T.,
  \& Hoffleit, D.\ 1995, The General Catalogue of Trigonometric
  Stellar Parallaxes (4th ed.; New Haven: Yale Univ. Obs.)

\bibitem[van de Kamp(1967)]{vandekamp} van de Kamp, P., 1967,
  Principles of Astrometry (San Francisco: W.~H.~Freeman and company)

\bibitem[Winters et al.(2011)]{Winters2011} Winters, J.~G., Henry,
  T.~J., Jao, W.-C., et al.\ 2011, \aj, 141, 21

\end{thebibliography}
\end{document}